\documentclass[twoside,11pt]{article}

\usepackage[ansinew]{inputenc}
\usepackage{graphicx}
\usepackage[lmargin=2.8cm, rmargin=2.8cm,bottom=3.15cm,top=3cm]{geometry}
\usepackage{amsfonts}
\usepackage{amsmath}
\usepackage{amsthm}
\usepackage{color}
\usepackage{amssymb}
\usepackage{url}
\usepackage{bm} 
\usepackage{dsfont} 

\usepackage{natbib}

\usepackage{sectsty}
\makeatletter\def\@seccntformat#1{\protect\makebox[0pt][r]{\csname the#1\endcsname\hspace{12pt}}}\makeatother

\usepackage{mathrsfs} 

\newcommand{\transpose}{^\top\! }

\newcommand{\inner}[2]{\left\langle{#1},{#2}\right\rangle}

\newcommand{\trace}{\mathrm{trace}}

\newcommand{\Rnn}{{\mathbb{R}^{n\times n}}}
\newcommand{\Rn}{{\mathbb{R}^n}}

\newcommand{\diag}{\mathrm{diag}}

\newcommand{\rank}{\operatorname{rank}}

\newcommand{\SHORTEN}[1]{}

\usepackage[autolinebreaks,useliterate]{mcode}

\newcommand{\calE}{E}
\newcommand{\calM}{\mathcal{M}}
\newcommand{\Rk}{\mathbb{R}^k}
\newcommand{\Rnm}{\mathbb{R}^{n\times m}}
\newcommand{\Rmm}{\mathbb{R}^{m\times m}}
\newcommand{\col}{\operatorname{col}}

\newcommand{\shorten}[1]{}

\begin{document}

\title{Manopt, a Matlab toolbox for optimization on manifolds}

\author{Nicolas Boumal\thanks{Corresponding author: nicolasboumal@gmail.com} \thanks{Department of Mathematical Engineering, Universit\'e catholique de Louvain, Louvain-la-Neuve, Belgium.} \and Bamdev Mishra\thanks{Department of Electrical Engineering and Computer Science, Universit\'e de Li\`ege, Li\`ege, Belgium.} \and P.-A.\ Absil\footnotemark[2] \and Rodolphe Sepulchre\footnotemark[3]}

\maketitle

\begin{abstract}

Optimization on manifolds is a rapidly developing  branch of nonlinear optimization. Its focus is on problems where the smooth geometry of the search space can be leveraged to design efficient numerical algorithms. In particular, optimization on manifolds is well-suited to deal with rank and orthogonality constraints. Such structured constraints appear pervasively in machine learning applications, including low-rank matrix completion, sensor network localization, camera network registration, independent component analysis, metric learning, dimensionality reduction and so on.

The Manopt toolbox, available at \url{www.manopt.org}, is a user-friendly, documented piece of software dedicated to simplify experimenting with state of the art Riemannian optimization algorithms. We aim particularly at reaching practitioners outside our field.



\vspace{3mm}


\end{abstract}

\section{Introduction}
Optimization on manifolds, or Riemannian optimization, is a fast growing research topic in the field of nonlinear optimization. Its purpose is to provide efficient numerical algorithms to solve optimization problems of the form
\begin{align}
	\min_{x\in\calM} f(x),
	\label{eq:minf}
\end{align}
where the search space $\calM$ is a smooth space: a differentiable manifold which can be endowed with a Riemannian structure. In a nutshell, this means $\calM$ can be linearized locally at each point $x$ as a tangent space $\mathrm{T}_x\calM$ and an inner product $\langle \cdot, \cdot \rangle_x$ which smoothly depends on $x$ is available on $\mathrm{T}_x\calM$. A number of smooth search spaces arise often in applications:

\shorten{
Such geometric structure in an optimization problem originates in mainly two ways. In some scenarios, problem~\eqref{eq:minf} is a constrained optimization problem for $x$ in $\calE$, a Euclidean space, such that $\calM$ is a smooth submanifold of $\calE$. For example, $\calE = \Rn$ and $\calM = \{x\in\Rn \colon x\transpose x = 1\}$. In other scenarios, problem~\eqref{eq:minf} comes from an unconstrained problem $\min_{u\in \calE} f(u)$  such that $f$ presents symmetries in the form of an equivalence relation $\sim$ over $\calE$: $u\sim v \Rightarrow f(u) = f(v)$. Then, $f$ is constant over equivalence classes $x = [u] = \{v\in\calE\colon u\sim v\}$ and descends as a well-defined function over the quotient space $\calM = E/\sim$.
}

\shorten{
Problem~\eqref{eq:minf} is defined based on $\calM$ being a set, regardless of any additional structure. In the context of optimization on manifolds though, it is precisely the Riemannian structure conferred to $\calM$ which will lead to efficient numerical methods to solve~\eqref{eq:minf}. When points of $\calM$ may be represented as points of a Euclidean space $\calE$, one often convenient way to equip $\calM$ with a Riemannian structure is to endow $\calE$ with a classical inner product (say, $\inner{X}{Y} = \trace(X\transpose Y)$ if $\calE = \Rnm$) and to structure $\calM$ as either a Riemannian submanifold or a Riemannian quotient manifold of $\calE$. 
}

\begin{itemize}

\item
The {\bf oblique manifold} $\calM = \{ X \in \Rnm \colon \diag(X\transpose X) = \mathds{1}_m \}$ is a product of spheres. That is, $X\in\calM$ if each column of $X$ has unit 2-norm in $\Rn$. \citet{absil2006jointdiag} show how independent component analysis can be cast on this manifold as non-orthogonal joint diagonalization. When furthermore it is only the product $Y = X\transpose X$ which matters, matrices of the form $QX$ are equivalent for all orthogonal $Q$. Quotienting out this equivalence relation yields the  {\bf fixed-rank elliptope} $\calM = \{ Y \in \Rmm : Y = Y\transpose \succeq 0, \rank(Y) = n, \diag(Y) = \mathds{1}_m \}$. \shorten{For increasing $n \geq 2$, this yields increasingly relaxed search spaces for max-cut, ultimately culminating in the acclaimed SDP relaxation of max-cut for $n = m$. \citet{journee2010low} show how to exploit this sequence of relaxed formulations of max-cut as Riemannian optimization problems to efficiently compute good cuts.} See the example below for application to the max-cut problem. The packing problem on the sphere, where one wishes to place $m$ points on the unit sphere in $\Rn$ such that the two closest points are as far apart as possible~\citep{dirr2007nonsmooth}, is another example of an optimization problem on the fixed-rank elliptope. \citet{grubisic2007lowrankcorrelation} optimize over this set to produce low-rank approximations of covariance matrices.

\item
The (compact) {\bf Stiefel manifold} is the Riemannian submanifold of orthonormal matrices, $\calM = \{ X\in\Rnm \colon X\transpose X = I_m \}$. \citet{amari1999natural} and \citet{theis2009soft} formulate versions of independent component analysis with dimensionality reduction as optimization over the Stiefel manifold.

\item
The {\bf Grassmann manifold} is the manifold $\calM = \{ \col(X) \colon X\in\Rnm_* \}$, where $\Rnm_*$ is the set of full-rank matrices in $\Rnm$ and $\col(X)$ denotes the subspace spanned by the columns of $X$. That is, $\col(X) \in \calM$ is a subspace of $\Rn$ of dimension $m$. \shorten{It is often given the geometry of a Riemannian quotient manifold of either $\Rnm_*$ or of the Stiefel manifold, where two matrices are equivalent if their columns span the same subspace.} Among other things, optimization over the Grassmann manifold proves useful in low-rank matrix completion, where it is observed that if one knows the column space spanned by the sought matrix, then completing the matrix according to a least squares criterion is easy~\citep{keshavan2010matrix,boumal2011rtrmc,balzano2010online}.

\item
The {\bf special orthogonal group} $\calM = \{ X \in \Rnn \colon X\transpose X = I_n \textrm{ and } \det(X) = 1 \}$ is the group of rotations, typically considered as a Riemannian submanifold of $\Rnn$. Optimization problems involving rotation matrices notably occur in robotics and computer vision, when estimating the attitude of vehicles or the pose of cameras~\citep{tron2009distributed,boumal2013MLE}.

\item
The set of {\bf fixed-rank matrices} $\calM = \{X \in \Rnm \colon \rank(X) = k\}$ admits a number of different Riemannian structures. \citet{vandereycken2013lowrank} proposes an embedded geometry for $\calM$ and exploits Riemannian optimization on that manifold to address the low-rank matrix completion problem. \citet{shalit2012online} use the same geometry to address similarity learning. \citet{mishra2012fixed} cover a number of quotient geometries for $\calM$ and similarly address low-rank matrix completion problems.

\item
The set of {\bf symmetric, positive semidefinite, fixed-rank matrices} is also a manifold, $\calM = \{ X \in\Rnn \colon X = X\transpose \succeq 0, \rank(X) = k \}$. \citet{meyer2011regression} exploit this to propose low-rank algorithms for metric learning. This space is tightly related to the space of {\bf Euclidean distance matrices} $X$ such that $X_{ij}$ is the squared distance between two fixed points $x_i, x_j \in \Rk$. \citet{mishra2011EDM} leverage this geometry to formulate efficient low-rank algorithms for Euclidean distance matrix completion.

\item
The {\bf fixed-rank spectrahedron} $\calM = \{ X \in \Rnn \colon X = X\transpose \succeq 0, \trace(X) = 1, \rank(X) = k \}$, without the rank constraint, is a convex set which can be used to solve relaxed (lifted) formulations of the sparse PCA problem. \citet{journee2010low} show how optimizing over the fixed-rank spectrahedron can lead to efficient algorithms for sparse PCA.

\end{itemize}

The rich geometry of Riemannian manifolds $\calM$ makes it possible to define gradients and Hessians of cost functions $f$, as well as systematic procedures (called \emph{retractions}) to move on the manifold starting at a point $x$, along a specified tangent direction at $x$. Those are sufficient ingredients to generalize standard nonlinear optimization methods such as gradient descent, conjugate-gradients, quasi-Newton, trust-regions, etc.

In a recent monograph, \citet{AMS08} lay down a mature framework to analyze problems of the form~\eqref{eq:minf} when $f$ is a smooth function, with a strong emphasis on building a theory that leads to efficient numerical algorithms. In particular, they describe the necessary ingredients to design first- and second-order algorithms on Riemannian manifolds in general. These algorithms come with convergence guarantees essentially matching those of the Euclidean counterparts they generalize. For example, the Riemannian trust-region method is known to converge globally toward critical points and to converge locally quadratically when the Hessian of $f$ is available. In many respects, this theory subsumes well-known results from an earlier paper by~\citet{edelman1998geometry}, which focused on problems of the form~\eqref{eq:minf} with $\calM$ either the set of orthonormal matrices (the Stiefel manifold) or the set of linear subspaces (the Grassmann manifold). 

The maturity of the theory of smooth Riemannian optimization, its widespread applicability and its excellent track record performance-wise prompted us to build the Manopt toolbox: a user-friendly piece of software to help researchers and practitioners experiment with these tools. Code and documentation are available at \url{www.manopt.org}.

%
%

\section{Architecture and features of Manopt}

The toolbox architecture is based on a separation of the manifolds, the solvers and the problem descriptions. For basic use, one only needs to pick a manifold from the library, describe the cost function (and possible derivatives) on this manifold and pass it on to a solver. Accompanying tools help the user in common tasks such as numerically checking whether the cost function agrees with its derivatives up to the appropriate order, approximating the Hessian based on the gradient of the cost, etc.

Manifolds in Manopt are represented as structures and are obtained by calling a factory. The manifold descriptions include projections on tangent spaces, retractions, helpers to convert Euclidean derivatives (gradient and Hessian) to Riemannian derivatives, etc. All the manifolds mentioned in the introduction work out of the box, and more can be added (shape space~\citep{ring2012optimization}, low-rank tensors~\citep{kressner2013tensors}, etc.). Cartesian products of known manifolds are supported too.

Solvers are functions in Manopt that implement generic Riemannian minimization algorithms. \shorten{All options have default values. }Solvers log standard information at each iteration and comply with standard stopping critera. Extra information can be logged via callbacks and, similarly, user-defined stopping criteria are allowed. Currently, Riemannian trust-regions (based on~\citep{genrtr}) and conjugate-gradients are implemented (with preconditioning), as well as steepest-descent and a couple derivative free schemes. More solvers can be added, with an outlook toward Riemannian BFGS~\citep{ring2012optimization}, stochastic gradients~\citep{bonnabel2013stochastic}, nonsmooth subgradients schemes~\citep{dirr2007nonsmooth}, etc. 

An optimization problem in Manopt is represented as a problem structure. The latter includes a field which contains a structure describing a manifold, as obtained from a factory. Additionally, the problem structure hosts function handles for the cost function $f$ and (possibly) its derivatives. An abstraction layer at the interface between the solvers and the problem description offers great flexibility in the cost function description. As the needs grow during the life-cycle of the toolbox and new ways of describing $f$ become necessary (subdifferentials, partial gradients, etc.), it will be sufficient to update this interface.


Computing $f(x)$ typically produces intermediate results which can be reused in order to compute the derivatives of $f$ at $x$. To prevent redundant computations, Manopt incorporates an (optional) caching system, which becomes useful when transiting from a proof-of-concept draft of the algorithm to a convincing implementation.

\section{Example: the maximum cut problem}


Given an undirected graph with $n$ nodes and weights $w_{ij} \geq 0$ on the edges such that $W \in \Rnn$ is the weighted adjacency matrix and $D\in\Rnn$ is the diagonal degree matrix with $D_{ii} = \sum_j w_{ij}$, the graph Laplacian is the positive semidefinite matrix $L = D-W$. The max-cut problem consists in building a partition $s \in \{+1, -1\}^n$ of the nodes in two classes such that $\frac{1}{4} s\transpose L s = \sum_{i<j} w_{ij} \frac{(s_i - s_j)^2}{4}$, i.e., the sum of the weights of the edges connecting the two classes, is maximum. Let $X = ss\transpose$. Then, max-cut is equivalent to:
\begin{align*}
	\max_{X \in \Rnn} & \trace(LX)/4 \\
	\textrm{s.t. } & X = X\transpose \succeq 0, \diag(X) = \mathds{1}_n \textrm{ and } \rank(X) = 1.
\end{align*}
\citet{goemans1995maxcut} proposed and analyzed the famous relaxation of this problem which consists in dropping the rank constraint, yielding a semidefinite program. Alternatively relaxing the rank constraint to be $\rank(X) \leq r$ for some $1 < r < n$ yields a tighter but nonconvex relaxation. \citet{journee2010low} observe that fixing the rank with the constraint $\rank(X) = r$ turns the search space into a smooth manifold, the fixed-rank elliptope, which can be optimized over using Riemannian optimization. In Manopt, simple code for this reads (with $Y\in\mathbb{R}^{n\times r}$ such that $X = YY\transpose$):
\begin{lstlisting}
% The problem structure hosts a manifold structure as well as function handles
% to define the cost function and its derivatives (here provided as Euclidean
% derivatives, which will be converted to their Riemannian equivalent).
problem.M = elliptopefactory(n, r);
problem.cost  = @(Y)    -trace(Y'*L*Y)/4;
problem.egrad = @(Y)    -(L*Y)/2;
problem.ehess = @(Y, U) -(L*U)/2;           % optional

% These diagnostics tools help make sure the gradient and Hessian are correct.
checkgradient(problem); pause;
checkhessian(problem);  pause;

% Minimize with trust-regions, a random initial guess and default options.
Y = trustregions(problem);
\end{lstlisting}
Randomly projecting $Y$ yields a cut: \mcode{s = sign(Y*randn(r, 1))}. The Manopt distribution includes advanced code for this example, where the caching functionalities are used to avoid redundant computations of the product $LY$ in the cost and the gradient, and the rank $r$ is increased gradually to obtain a global solution of the max-cut SDP (and hence a formal upperbound), following the procedure in~\citep{journee2010low}.

%
%
%
%
%
%

\clearpage

\section*{Acknowledgments}
NB and BM are research fellows with the FNRS (aspirants). This paper presents research results of the Belgian Network DYSCO (Dynamical Systems, Control, and Optimization), funded by the Interuniversity Attraction Poles Programme initiated by the Belgian Science Policy Office. It was financially supported by the Belgian FRFC.

\bibliographystyle{authordate1}
\bibliography{boumal}

\end{document}